\documentclass{revtex4-2}

\usepackage{amsmath,amsfonts,amssymb,latexsym,bm}
\usepackage{graphicx}
\usepackage{algorithm}
\usepackage{algorithmic} 
\usepackage{subcaption}
\usepackage[utf8]{inputenc}

\def\hamilt{\hat H}
\def\s{\sigma}
\def\b{\beta}
\def\r{\hat{\rho}}

\def\pauz{\hat{\sigma}^z}
\def\paux{\hat{\sigma}^x}

\DeclareMathOperator\arctanh{arctanh}
\newcommand{\ket}[1]{|#1\rangle}
\newcommand{\Mij}[2]{M_{#1 \rightarrow #2}(\vec \sigma_{#2})}
\newcommand{\MUij}[2]{\mu_{#1 \rightarrow #2}(\vec \sigma_{#1})}
\newcommand{\tr}[1]{\text{Tr}[#1]}

\newcommand{\bfs}{\bm{\sigma}}
\newcommand{\expectation}[1]{\left\langle #1 \right\rangle}

\begin{document}
\title{Efficient inference in the transverse field Ising model}
\author{E. Dom\'{\i}nguez}
\email{eduardo.dominguezvazquez@donders.ru.nl}
\affiliation{Donders Institute for Brain, Cognition and Behavior. Radboud University. The Netherlands}
\author{H.J. Kappen}
\email{b.kappen@science.ru.nl}
\affiliation{Donders Institute for Brain, Cognition and Behavior. Radboud University. The Netherlands}

\date{\today}

\begin{abstract}
\noindent In this paper we introduce an approximate method to solve the quantum cavity equations for transverse field Ising models. The method relies on a projective approximation of the exact cavity distributions of imaginary time trajectories (paths). A key feature, novel in the context of similar algorithms, is the explicit separation of the classical and quantum parts of the distributions. Numerical simulations show accurate results in comparison with the sampled solution of the cavity equations, the exact diagonalization of the Hamiltonian (when possible) and other approximate inference methods in the literature. The computational complexity of this new algorithm scales linearly with the connectivity of the underlying lattice, enabling the study of highly connected networks, as the ones often encountered in quantum machine learning problems.

\end{abstract}
\maketitle

\section{Introduction}

Many relevant systems in statistical mechanics, combinatorial optimization and quantum information
theory can be mapped to realizations of the so-called transverse field Ising model (TFIM) \cite{Dutta15,Sachdev}. Notable examples include Kitaev 1D chains \cite{kitaev2010topological,1DIsingKitaev}, many NP-complete problems such as 3SAT \cite{Ising_NP} and the foundations of the whole field of adiabatic quantum computation \cite{Nishimori,Farhi2000QuantumCB,dwave}. Given its numerous applications, the study of the properties of this lattice spin model has attracted considerable attention in the last decades. Thanks to that, much have been learned about the nature of quantum phase transitions, especially for the 1D case. However, despite all efforts, doing inference (i.e. computing observables) remains very hard in the general case. The problem stems, as usual, from the exponential increase in size of the Hilbert space with the number of variables. There are two potential ways around the dimensionality curse. One is using a quantum computer to directly implement the required Hamiltonian and let Nature deal with the Hilbert space. Unfortunately, the development of a general purpose quantum computer seems to be at least some decades in the future and near term devices are challenged by decoherence effects and thermal fluctuations. The other option is to make use of suitable approximations and make computations in a classical device.  

The quantum cavity method (QC), introduced in \cite{laumann,florent}, enables a reduction of the complexity of the inference tasks for a TFIM.  It combines an imaginary-time path integral expansion of the density matrix with the cavity formalism, well known in the study of classical disordered systems \cite{cavity_method_original}.  As in the conventional cavity approach, QC establishes that all the information of the system is encoded in the set of single variable \textit{cavity} distributions for interaction networks lacking short loops. This is not true for general topologies but QC remains a useful mean field approximation in that scenario. 

The set of coupled equations that determine the cavity distributions can be, at least in principle, solved by an iterative procedure, formally equivalent to the message passing algorithm known as Belief Propagation (BP) \cite{yedidia}. In practice, the implementation cost is similar to a population dynamics algorithm \cite{mezard2001bethe}, with the important difference that in the present case the sampled variables are trajectories and not real values. There are problems for which the inference calculations must be repeated many times, e.g. for the learning loop of a quantum Boltzmann machine \cite{QBM_Kappen,QBM_Amin}. Therefore, it is highly desirable to develop time- and memory-efficient (albeit approximated) methods to solve the QC equations. This is the main purpose of the present work.
By focusing of the structure of the QC equations and the properties of the cavity distributions we will be able to derive a fast, precise and flexible solution to the inference problem for the TFIM. To put our analysis in context, we will consider the predictions of other previously developed inference methods. Moreover, we will show how these approximations fit into the QC formalism. 

Following the introduction of QC, approximated solution schemes were devised to further reduce its computational burden. Already in \cite{florent}, the static approximation was proposed as a cheaper way to obtain qualitatively correct results. The static numerical predictions, however, are far from ideal. Of special interest to us is the projective procedure employed in \cite{Dimitrova}, by which the information in the (cavity) distribution is compressed to just a couple effective field parameters. Remarkably, the projected cavity method (PCM), developed therein, is quite precise in the numerical estimation of the observables and the critical temperature of sparse ferromagnetic lattices. Two drawbacks of PCM are the disadvantageous scaling (exponential) of the running time with the system connectivity and that it is specially fine tuned for longitudinal observables, somewhat at the expense of less transversal precision. Other approximations considered in \cite{Dimitrova} are based on more qualitative mean field arguments. This is the case of the so-called naive mean field and the cavity mean field (CMF) methods. The CMF proposal is simpler and faster than PCM but, as we will see, it does not reduce to the cavity solution in the classical limit of no tranverse field. This should be a minimal requirement to all approximate solutions of the quantum cavity equations that aim at being useful for a wide range of values of the external parameters.

An inference method applicable to the TFIM is the quantum Cluster Variation Method and the related algorithm  called quantum Belief Propagation (qBP) \cite{qcvm,Tanaka_2009}. At first glance, the standard derivation based on the minimization of a variational Bethe free energy functional is somewhat removed from the QC formalism. However, we can show that it is possible to establish a connection and consider qBP as an approximate solution scheme of the QC equations.

The rest of the paper is organized as follows. In section \ref{sec:QC}, we give a precise definition of the TFIM and review the quantum cavity approach. We recast the exact cavity equations in a form that is suitable for subsequent approximate treatments. In section \ref{sec:approx_QC} we will introduce 
the main result of this work, an inference algorithm that we call quantum cavity mean field (qCMF), and discuss
the connections with other inference approximations. Numerical simulations in different scenarios and topologies
are discussed in section \ref{sec:results}. Finally, in section \ref{sec:conclusion} we provide an overview of the results, some concluding remarks and perspectives for future developments.

\section{Quantum cavity solution of the TFIM}
\label{sec:QC}
In this section we present a summary of the quantum cavity formalism as applied to the transverse field Ising model on a diluted graph. We will introduce the basic Hamiltonian of the system, the corresponding density matrix and sketch the mapping that turns the quantum statistical problem into a classical probability distribution. We skip the detailed derivation and refer the interested reader to the original papers \cite{florent,laumann}.

The TFIM is defined by the following Hamiltonian: 
\begin{equation}
\hamilt = - \sum_{(ij)} J_{ij} \pauz_i \pauz_j - \sum_{i} h_i \pauz_i -  \sum_{i} B_i\paux_i
\label{eqn:ising_hamiltonian}
\end{equation}
The transverse fields $B_i$ introduce non-commuting terms to an otherwise classical (diagonal) Hamiltonian. In the context of the cavity method, the interaction topology specified by the
symmetric matrix $J_{ij}$ corresponds to a diluted, tree-like network. Realizations
of such geometries include Cayley trees as well as random regular and Erdos-Renyi graphs.
One can also apply the cavity method in other situations, such as in finite dimensional regular lattices. In that case, instead of producing an exact solution, it is interpreted as an structured mean field approximation.

\begin{figure}[h]
 \centering
  \includegraphics[width=0.4\textwidth,keepaspectratio=true]{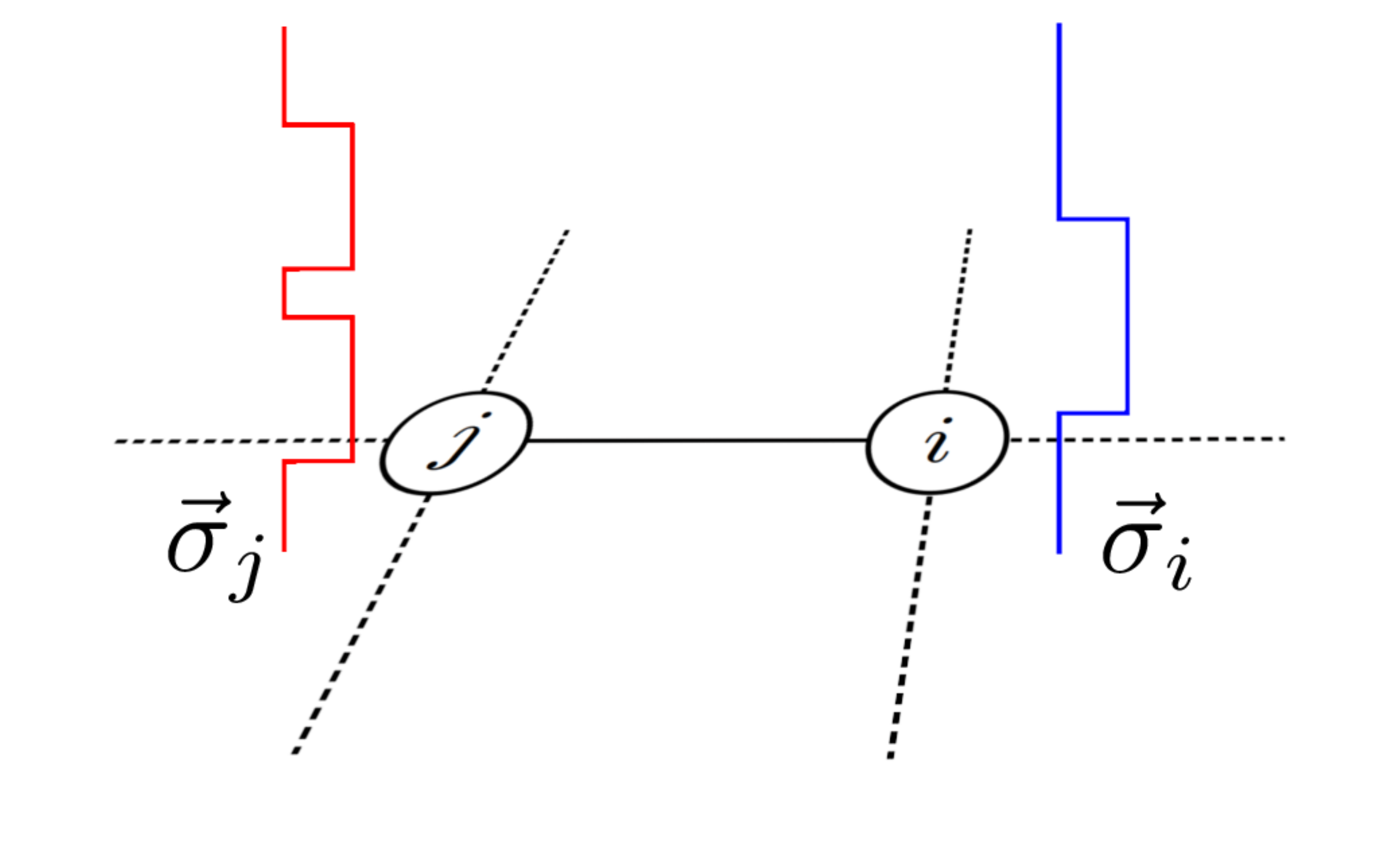}
 \caption{\label{fig:imaginary_time}
 The classical variables $\vec \s_i$ are
 stepwise functions of the (discrete or continuous) imaginary time coordinate.}
\end{figure}

The quantum cavity method is based on a Suzuki-Trotter
expansion of the density matrix $\r = \frac{1}{Z}\exp -\beta \hamilt$ \cite{suzuki1976}. This procedure reveals a mapping between the quantum system defined by \eqref{eqn:ising_hamiltonian} and an equivalent classical model with
an extra dimension (the so-called imaginary time coordinate). The classical
equivalent has the same interaction topology as the original, with trajectories (paths) replacing the quantum spins in the nodes of the network structure.
The statistics of the new system will be given by the joint probability distribution $\rho(\bm{\sigma})$ of all trajectories $\bm{\sigma}= (\vec{\s}_1,\ldots,\vec{\s}_N)$. 
Each path, denoted $\vec{\s}_i$, is a piecewise constant function
taking the values $\pm 1$ alternatively in the imaginary time direction, see Fig. (\ref{fig:imaginary_time}).
A precise definition of $\rho(\bm{\sigma})$ is given in equations \eqref{eqn:ST_density} to \eqref{eqn:ST_w}:
\begin{align}
\label{eqn:ST_density}
 \rho(\bfs)  &= \dfrac{1}{Z}\exp\left[ -\beta E_z(\bfs)\right] \prod_i w(\vec\s_i,B_i)\\
 \label{eqn:ST_Ez}
 E_z(\bfs) &= - \sum_{(ij)} J_{ij} \vec \s_i \cdot \vec\s_j - \sum_{i} h_i m(\vec\s_i)\\
 w(\vec\s_i,B_i) &= \prod_{\tau=1}^{N_s} \langle \s_i^\tau | \exp\left[\b \dfrac{B_i}{N_s} \paux \right] |  \s_i^{\tau + 1}\rangle 
 \label{eqn:ST_w}
 \end{align}
In equation \eqref{eqn:ST_Ez} we have defined $m(\vec\s_i) = \frac{1}{N_s} \sum_{\tau} \s_i^\tau$ and $\vec \s_i \cdot \vec\s_j = \frac{1}{N_s} \sum_{\tau} \s_i^\tau \s_j^\tau$, where $N_s$ is the number of Suzuki-Trotter slices used
in the expansion. The $\tau$ index is the imaginary time coordinate, taking values in the range $[1,\ldots,N_s]$. We will, however, normally work in the continuous time limit $N_s \rightarrow \infty$. Then $\tau$ becomes a continuous parameter and the previous sums become definite integrals in the interval $[0,1]$. Notice that the energy function $E_z(\bfs)$ does not contain interactions between different Trotter slices. Moreover, the external fields $h_i$ act homogeneously in the 
imaginary time direction. Different slices are coupled only by the $w(\vec \s_i,B_i)$ factors. The form of these $w$ factors imply that in the classical regime, $B_i \rightarrow 0$, the trajectories with non-zero probability are only those which are constant in the imaginary time direction. It is also easy to
see in the definition of $w(\vec\s_i,B_i)$ that this interaction represents a one-dimensional ferromagnetic chain in the imaginary time direction.

The quantum-classical mapping is completed once we specify the correspondence of quantum operators representing physical observables with functions of the spin trajectories: $\hat F \rightarrow F(\bfs)$.
The quantum average $\langle \hat F \rangle = \tr{\r \hat F}$ becomes a classical average 
$\sum_{\bfs} F(\bfs) \rho(\bfs)$. For example, to compute $m^z_i = \langle \pauz_i\rangle$
one has to average $m(\vec \s_i)$ over the marginal $\rho_i(\vec\s_i)$:
\begin{equation}
 m^z_i = \sum_{\vec\s_i} m(\vec\s_i) \rho_i(\vec\s_i)
 \label{eqn:mz_mag}
\end{equation}
See Appendix \eqref{app:observables} for the definitions of other observables of interest such as the $m_i^x$ magnetization
and pairwise correlations. The important point to note here is that these
quantities typically rely on the computation of marginals of the joint density
$\rho(\bfs)$, that is, on the distributions of small subsets of trajectories.
The inference of such marginals is computationally unfeasible, much like in the original quantum system, where the exponentially growing size of the Hilbert space forbids any attempts of computing partial traces of the density matrix. At this point is where the cavity method comes into action.

The cavity approach exploits the structure of the interaction network to simplify the computation of local marginals such as $\rho_i(\vec\s_i)$ and $\rho_{ij}(\vec\s_i, \vec\s_j)$. The functional form of the resulting local distributions include a Boltzmann-like term with the interactions appearing in the original Hamiltonian. The effect of the neighbors is included by extra multiplicative factors. The single spin (trajectory) distribution has the form:
\begin{equation}
 \rho_i(\vec \s_i) = \dfrac{1}{Z_i} w(\vec \s_i,B_i)
 \exp\left[\beta h_i m(\vec \s_i)\right] 
 \prod_{k \in \partial i} \Mij{k}{i}
 \label{eqn:single_site_distribution}
\end{equation}
here $\partial i$ represents the set of spins that are neighbors of spin $i$.
For interacting pairs we have a similar structure: 
\begin{equation}
 \rho_{ij}(\vec \s_i,\vec \s_j) = \dfrac{1}{Z_{ij}} 
 \exp\left[\beta J_{ij} \vec\s_i\cdot\vec\s_j \right] \MUij{i}{j} \MUij{j}{i}
 \label{eqn:pair_distribution}
\end{equation}

The $\mu$ and $M$ factors above are called \textit{cavity} distributions. Any given $\MUij{i}{j}$ is interpreted as the probability distribution of the variable $\vec\s_i$ in a system where the interaction with site $j$ is removed. Contrastingly, although $\Mij{k}{i}$ is positive for all $\vec \s_i$ and can be normalized to a probability distribution, its physical meaning is not so clear in this quantum scenario. A reader familiar with the classical Belief Propagation will identify \eqref{eqn:single_site_distribution} and \eqref{eqn:pair_distribution} as the \textit{belief} distributions and the $M$ and $\mu$ functions will correspond to messages in a factor graph \cite{yedidia}. 

The cavity distributions $\mu$  and  $M$ must take values such that the consistency of \eqref{eqn:single_site_distribution} and \eqref{eqn:pair_distribution} holds. The probability function of any pair must be consistent with each single trajectory distribution:
\begin{equation}
 \rho_i(\vec \s_i) = \sum_{\vec \s_j}\rho_{ij}(\vec \s_i,\vec \s_j)
 \label{eqn:consistency_bp}
\end{equation}
The constraints introduced by \eqref{eqn:consistency_bp} give the relation between the $\mu$ and $M$ distributions:
\begin{eqnarray}
\label{eqn:message_passing_mu}
 \MUij{i}{j} &=& \dfrac{1}{Z_{i\rightarrow j}^\mu}  w(\vec \s_i,B_i)
 \exp\left[\beta h_i m(\vec \s_i)\right] 
 \prod_{k \in \partial i\setminus j} \Mij{k}{i}\\ 
 \Mij{k}{i} &=& \dfrac{1}{Z_{k\rightarrow i}^M} 
 \sum_{\vec\s_k}\exp\left[\beta J_{ik} \vec\s_i\cdot\vec\s_k \right] \MUij{k}{i}
 \label{eqn:message_passing_M}
\end{eqnarray}

An exact solution of the system of equations given by \eqref{eqn:message_passing_mu} and \eqref{eqn:message_passing_M} is technically
unfeasible in the general case, due to the characteristics of the sample space.
 One possibility is to implement a numerical solution as described in \cite{florent}. The basic idea is to represent the $\mu$ distributions with sampled estimates, that is, each distribution becomes a weighted collection of trajectories. 

In the next section we will describe alternative approaches. We will explore approximations where the complexity of the distributions is reduced with the aid of mean field arguments. Only a reduced number of parameters will be needed to encapsule the statistical properties of the cavity distributions.

As a final note in this summary, it is worth mentioning that it is not strictly necessary to use both $\mu$ and $M$ distributions. One can perfectly choose one type at convenience and write everything in terms of that. In fact, previous contributions privilege the use of $\mu$ distributions, probably due to the clearer physical interpretation. We think, however, that adopting a combined treatment provides further insights and flexibility regarding where and how approximations can be done. For completeness, we write down the self-consistent
equations that are obtained after eliminating the $M$ distributions:
\begin{equation}
\label{eqn:message_passing_mu_no_M}
 \MUij{i}{j}\propto w(\vec \s_i,B_i)
 \exp\left[\beta h_i m(\vec \s_i)\right] 
 \prod_{k \in \partial i\setminus j} \sum_{\vec\s_k}\exp\left[\beta J_{ik} \vec\s_i\cdot\vec\s_k \right] \MUij{k}{i}.
\end{equation}
and the ones resulting from eliminating $\mu$:
\begin{equation}
\label{eqn:message_passing_mu_no_mu}
 \Mij{k}{i} \propto 
  \sum_{\vec\s_k}\exp\left[\beta J_{ik} \vec\s_i\cdot\vec\s_k + \b h_k m(\vec\s_k)\right]  w(\vec \s_k,B_k)  \prod_{l \in \partial k\setminus i} \Mij{l}{k}
\end{equation}

\section{Approximate solutions of the cavity equations}
\label{sec:approx_QC}
Approximate solutions of the cavity equations are usually of a projective type. 
This means that one looks for a solution within a tractable subspace (e.g. of
factorized distributions). The main problem with this approach is that
equations \eqref{eqn:message_passing_mu} and \eqref{eqn:message_passing_M} do
not preserve the simplified structure in general. For example, if a factorized
(in the imaginary time direction) $\mu$ distribution is plugged in the RHS of 
\eqref{eqn:message_passing_M}, the resulting $M$ will not be factorized in the
same way. One needs to project said $M$ back to the easier subspace. Different projections lead to approximations with varying performance. Below we mention some possibilities already explored in the literature, as this will provide the necessary context for our own contribution. 

The static approximation \cite{florent} essentially amounts to introduce the ansatz $ \MUij{i}{j} = \mu_{i\rightarrow j} (m(\vec \s_i))$. The probability is projected over surfaces of constant $m(\vec\s_i)$. In other words, all trajectories with the same $m(\vec\s_i)$ get the same probability. Interestingly, the ansatz is correct in the classical limit, where trajectories are constant in the imaginary time direction. However, in the quantum regime this mean field treatment of the intrisically one dimensional structure of $\MUij{i}{j}$ [due to the $w$ factor in it, see equation \eqref{eqn:message_passing_mu_no_M}] leads to inaccurate numerical results.

A significantly more precise treatment is the projected cavity mapping (PCM) of \cite{Dimitrova, Ioffe2010a}. The idea in this case is to use a projection of the form $ \MUij{i}{j} = w(\vec \s_i,B_i) \exp\left[\beta h_{i\rightarrow j} m(\vec \s_i)\right]$. One can put this ansatz on the RHS of 
\eqref{eqn:message_passing_mu_no_M} for the distributions $\MUij{k}{i}$ and compute $h_{i\rightarrow j}$ on the LHS such that both parts of the equation give the same $m^z_i$ magnetization. The computation in the RHS is equivalent to
determining the $m^z_i$ magnetization of spin $i$ with the cavity Hamiltonian:
\begin{equation}
 \hat H_{\partial i\setminus j} = - B_i\paux_i -  h_i \pauz_i - \sum_{k\in \partial i\setminus j} \left[ J_{ik}\pauz_i  \pauz_k  +  h_{k\rightarrow i}
 \pauz_k + B_k \paux_k \right]
 \label{eqn:cavity_hamiltonian_PCM}
\end{equation}
In the LHS, the equivalent Hamiltonian is just $- B_i \paux_i -h_{i\rightarrow j} \pauz_i$. Equation \eqref{eqn:cavity_hamiltonian_PCM} reveals the fundamental
disadvantage of PCM, namely, the scaling of the computation time with the number of neighbors. This is a serious issue if one is interested in using the PCM to compute the approximate statistics of a model with high connectivity. 

In the same paper \cite{Dimitrova}, the authors propose what they call a cavity mean field (CMF) approximation to compute the single site statistics. It is a less refined approximation where the cavity Hamiltonian is defined as the single particle operator:
\begin{equation}
 \hat H_{i\setminus j} = - B_i\paux_i - \pauz_i \left[  h_i  + \sum_{k\in \partial i\setminus j} J_{ik} \langle \pauz_k\rangle_{k\setminus i} \right]
 \label{eqn:cavity_hamiltonian_CMF}
\end{equation}
In this case the projected cavity field $h_{i\rightarrow j}$ is simply  the expression inside brackets in equation \eqref{eqn:cavity_hamiltonian_CMF}.
The quantity $\langle \pauz_k\rangle_{k\setminus i}$ is computed in the cavity construction corresponding to spin $k$, that is, using the Hamiltonian $\hat H_{k\setminus i}$. This gives a recursive relation for the cavity fields:
\begin{equation}
 h_{i\rightarrow j} = h_i +  \sum_{k\in \partial i\setminus j} J_{ik} \dfrac{h_{k\rightarrow i}}{\sqrt{B_k^2 + h_{k\rightarrow i}^2}}
 \tanh \b \sqrt{B_k^2 + h_{k\rightarrow i}^2}
\end{equation}
The computational cost of this approximation scales much better (linearly) with the connectivity of the spins. The simplification comes at a cost: it does not provide the correct classical limit when $B\rightarrow 0$. In effect, if we take this limit, we get the iterative equations:
\begin{equation}
 h_{i\rightarrow j} = h_i + \sum_{k \in \partial i\setminus j} J_{ik} \tanh \beta h_{k\rightarrow i}
\end{equation}
whereas the correct classical relations for the cavity fields are \cite{mezard2001bethe}:
\begin{equation}
 h_{i\rightarrow j} = h_i + \sum_{k \in \partial i\setminus j}\frac{1}{\beta} \arctanh\left[ \tanh \beta J_{ik} \tanh \beta h_{k\rightarrow i} \right]
\end{equation}
Also, it is not clear from the original derivation what is the relation between the CMF ansatz and the formalism of trajectory distributions. To make this connection clearer we can go back to equation \eqref{eqn:message_passing_M} and approximate the average over $\vec\s_k$ by making:
\begin{equation}
\Mij{k}{i} \propto  \exp\left[\beta J_{ik} m(\vec\s_i) m_{k\rightarrow i}\right]
 \label{eqn:message_passing_M_CMF}
\end{equation}
This can be interpreted as substituting the trajectory $\vec\s_k$ in the exponent by its average value $m_{k\rightarrow i} = \langle m(\vec\s_k)\rangle_{\mu_{k\rightarrow i}}$, computed with the cavity distribution $\mu_{k\rightarrow i}$. After using
this ansatz in equation \eqref{eqn:message_passing_mu}, the effective field $h_{i\rightarrow j}$  is identical to the CMF result.  
Although the CMF can be obtained without making any contact with the trajectory formalism, in doing so we can readily
find pointers for improvements. This is an attractive goal due to the convenient scaling properties of this algorithm. We note in passing that the naive mean field approximation of \cite{Dimitrova} can also be obtained from the exact cavity equations
if we make the approximation $\Mij{k}{i} \propto  \exp\left[\beta J_{ik} m(\vec\sigma_i) m_k \right]$ where $m_k$ is computed with complete distribution $\rho_k(\vec\s_k)$ instead of the cavity one $\MUij{k}{i}$. The fixed point equations for the magnetizations $m_i$ turn out to be the same in both formalisms. 

Another inference method we want to discuss is a quantum version of 
the Belief Propagation algorithm (qBP) \cite{}.
The usual derivation follows a minimization of the Bethe free energy
of the system in a restricted subspace of compatible local density matrices
$\r_i$ and $\r_{ij}$. While searching for effective projection strategies, we have found that it is also possible to obtain the qBP algorithm from the full quantum cavity equations. To see this, one can use the $M$ distributions and project to the exponential form $M_{k\rightarrow i}(\vec\s_k)\propto \exp\left[\b u_{k\rightarrow i}m(\vec\s_k)\right]$. This time, however, the value of $u_{k\rightarrow i}$ will not be fixed  by projecting in \eqref{eqn:message_passing_M} or \eqref{eqn:message_passing_mu_no_mu}. Instead, just as in
the variational derivation of qBP, the effective field $u_{k\rightarrow i}$ is fixed
by forcing the consistency relation: 
\begin{equation}
 \sum_{\vec \s_i} m(\vec \s_i) \rho_i(\vec \s_i) = \sum_{\vec \s_i} m(\vec \s_i) \sum_{\vec \s_k}\rho_{ik}(\vec \s_i,\vec \s_k)
 \label{eqn:consistency_bp_mz}
\end{equation}
One can show that the effective fields $u_{k\rightarrow i}$ obey the same recurrence equations as in qBP. 
This quantum extension of BP has an important structural difference with the classical one. For classical BP, in the computation of $u_{k\rightarrow i}$ only the set of effective fields  $\{u_{l\rightarrow k}\}$  is needed (with $l$ running over the neighbours of $k$ except for spin $i$). This means that a change in $u_{k\rightarrow i}$ would never have an effect on itself if the network is a tree. For classical trees, information flows always in loopless paths. In the quantum case, however, the situation is quite different. Equation \eqref{eqn:consistency_bp_mz} uses also information from neighbors of spin $i$ to determine $u_{k\rightarrow i}$. It is not hard to see that in this case there will be feedback loops for the flow of information. Numerical simulations discussed in section \ref{sec:results} show that this induced loopiness has consequences for the behaviour of the algorithm.

\subsection{The quantum cavity mean field algorithm}
In order to account for the  inaccuracy of CMF in the small $B$ regime,  we should look more closely to the behaviour of the classical and quantum parts of the trajectory distributions. Numerical studies for the static approximation \cite{florent} already suggest that the constant trajectories with $\s(\tau)= 1 (-1) \;\;\forall \tau$ bear a significant fraction of the probability weight for small and even moderate transverse field $B$ values. Separating the classical and quantum parts of the $\mu$ distributions before making any projective approximation is therefore an idea worth exploring. Using this insight we introduce below a \textit{quantum} Cavity Mean Field (qCMF) algorithm.

Let $I_c[\vec\s , s]$ be the indicator function that is different from zero only for the classical trajectories $\vec\s\equiv s$, with $s = \pm 1$:
\begin{equation}
 I_c[\vec\s , s]=\left\{
 \begin{array}{ll}
  1 & \;\;\s(\tau) = s\;\; \forall \tau \\
  0 & \;\;\text{otherwise}
 \end{array}
\right.
\end{equation}
and let $I_{q}[\vec\s]$ be the complement of $I_c$. We can then separate:
\begin{equation}
 \MUij{k}{i} = \MUij{k}{i} I_{q}[\vec\s_k] + \sum_{s = \pm 1} \MUij{k}{i} I_c[\vec\s_k , s]
 \label{eqn:quantum_classical_separation_MU}
\end{equation}
where the first and second term correspond to quantum and classical trajectories, respectively.
For convenience we will denote $c_{ki}(s) = \mu_{k\rightarrow i}(\vec\s_k\equiv s)$ the probability weight of the constant trajectories. It is clear that $q_{ki}  = 1-\sum_{s=\pm 1} c_{ki}(s)$ equals the 
trace of the quantum part of $\mu_{k\rightarrow i}$. With this definition we can renormalize the distribution restricted to the quantum trajectories by writing:
\begin{equation}
 \mu^q_{k\rightarrow i}(\vec \s_k) = \dfrac{1}{q_{ki}} \mu_{k\rightarrow i}(\vec \s_k) I_{q}[\vec\s_k]
\end{equation}
and obtain finally:

\begin{equation}
 \MUij{k}{i} = q_{ki}\mu^q_{k\rightarrow i}(\vec\s_k) + \sum_{s=\pm 1} c_{ki}(s) I_c[\vec\s_k , s]
 \label{eqn:split_mu}
\end{equation}
The key idea to outperform CMF is to use \eqref{eqn:split_mu} back in \eqref{eqn:message_passing_M} and project only the quantum part:
\begin{equation}
 \Mij{k}{i} \propto q_{ki} \exp\left[\beta J_{ik} m(\vec\s_i) m_{k\rightarrow i}^q \right] + \sum_{s=\pm 1} c_{ki}(s) \exp\left[\beta J_{ik} m(\vec\s_i)s\right]
 \label{eqn:message_passing_M_qCMF}
\end{equation}
The cavity magnetization $m_{k\rightarrow i}^q$ is computed using only the quantum part $\mu_{k\rightarrow i}^q$.

The functional form of \eqref{eqn:message_passing_M_qCMF} is more involved than just the single exponential term used in CMF but, since $\Mij{k}{i}$ is a function of $m(\vec\s_i)$ and not of all the details of the trajectory $\vec\s_i$, the problem is still tractable.
With the projected $\Mij{k}{i}$, computed for all $k\in \partial i\setminus j$, we can find the parameters $c_{ij}(s)$ and $m^q_{i\rightarrow j}$ for $\MUij{i}{j}$ and generate a new $\Mij{i}{j}$ with an expression similar to \eqref{eqn:message_passing_M_qCMF} [see Appendix \eqref{app:details_qCMF}]. This iteration is continued until a stationary value is reached for all the parameters.  Algorithm \eqref{alg:qCMF} provides a summary of the process.

\begin{algorithm}[H]
\caption{Quantum cavity mean field (qCMF)}\label{alg:qCMF}
\begin{algorithmic}[1]
\STATE Initialize the parameters $c_{ki}(s)$ and $m^q_{k\rightarrow i}$ for every $M_{k\rightarrow i}$ in the system.
\STATE For all pairs $(k,i)$ of connected spins compute the function $M_{k\rightarrow i}(m) = q_{ki} \exp\left[\beta J_{ik} m\, m_{k\rightarrow i}^q \right] + \sum_{s=\pm 1} c_{ki}(s) \exp\left[\beta J_{ik} m\,s\right]$ in the interval $m\in [-1,1]$. 
\STATE Using the $M_{k\rightarrow i}(m)$ functions, compute $\mu_{i\rightarrow j}(m)$ [see Appendix \eqref{app:details_qCMF} for details]. The cost of this step increases linearly with the connectivity.
\STATE From $\mu_{i\rightarrow j}(m)$ extract $c_{ij}(s)$ and compute $m^q_{i\rightarrow j}$.
\STATE Repeat from step 2 until convergence of all $c_{ki}(s)$ and $m^q_{k\rightarrow i}$.
\STATE Compute observables using the single spin and pair distributions given by \eqref{eqn:single_site_distribution}, \eqref{eqn:pair_distribution}.
\end{algorithmic}
\end{algorithm}

We show that the algorithm we are proposing here has the correct classical limit in Appendix \ref{subsec:classical_limit}. It is important to stress that the scaling of the computational cost of computing the $\mu$ distributions is linear in the number of spins in the cavity construction. 

The derivation we are presenting makes use of an ad-hoc projection of the quantum part of the $M$ distributions
to a tractable form. It would have been desirable to obtain the same result in a variational way, by minimizing
a suitable Bethe free energy. Unfortunately, we have not been able to find such derivation. What can be done to 
approximately compute the (Bethe) free energy of the system with the qCMF is to write the local distributions \eqref{eqn:single_site_distribution} and \eqref{eqn:pair_distribution} using the qCMF approximate
results. With these distributions it is possible  to compute the local site and pair free energies required for the Bethe approximation. For more details about the estimation of the Bethe free energy 
the reader can check Appendix \eqref{subsec:bethe_free_energy}.

\section{Results and discussion}
\label{sec:results}

In this section we present a numerical comparison of the quality
of the different approximated treatments of the quantum cavity method. We compare the static approximation (SA), the projected cavity method (PCM) and its efficient approximation (CMF), our proposed quantum cavity mean field (qCMF) and the quantum belief propagation (qBP). 

\begin{figure}[h]
 \centering
 \includegraphics[width=0.3\textwidth,keepaspectratio=true]{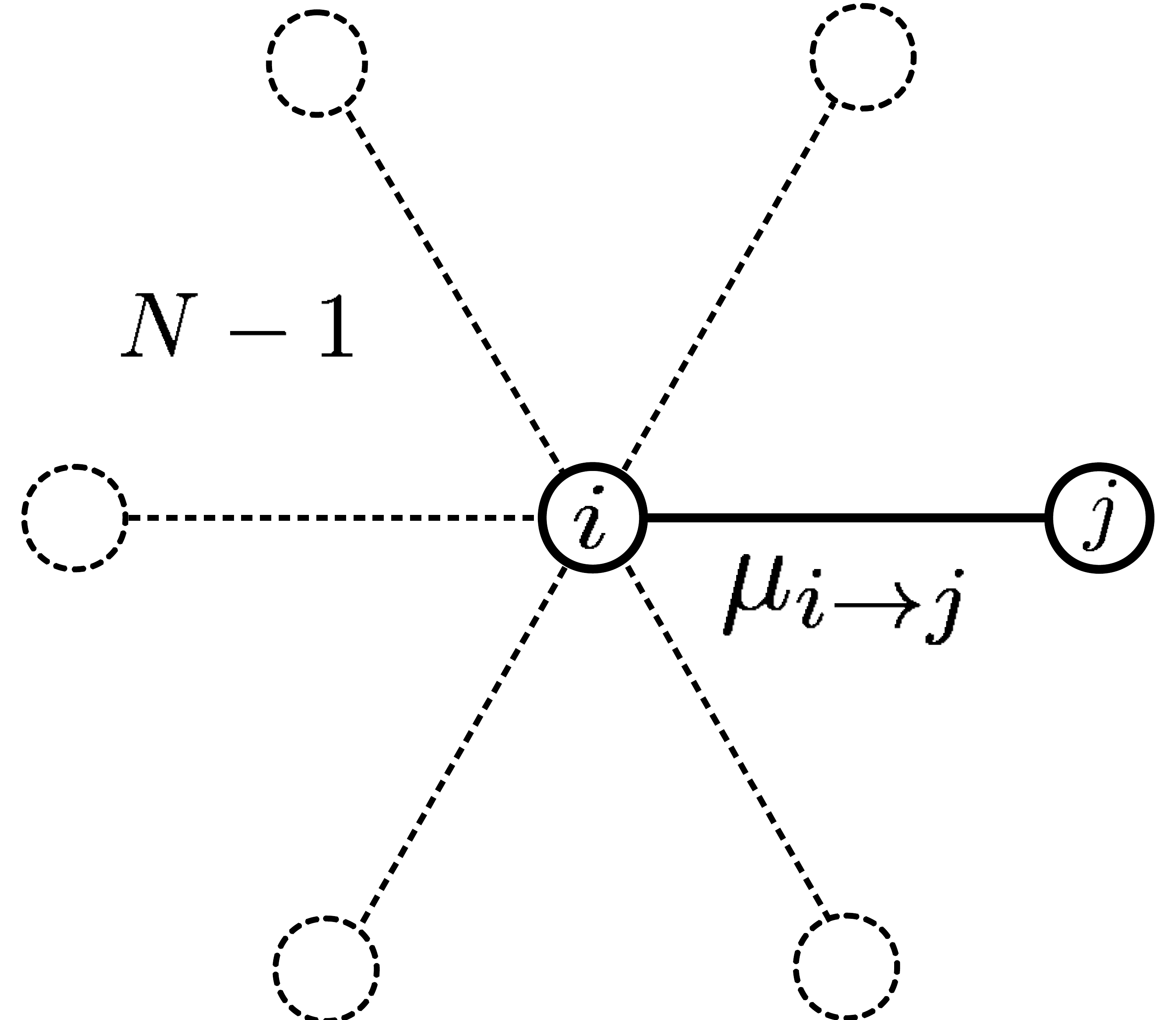}
 \caption{ \label{fig:star_model_diagram}Star model. We compute the cavity distribution $\MUij{i}{j}$ with the PCM and the qCMF approximation and compare the observables of spin $j$ in the edge for each case. The external fields $h$ and $B$ are the same for all spins. We study the quality of these approximations for different system sizes $N\in[3,12]$ using as
 a reference the computation of the exact density matrix, see figure \eqref{fig:star_model}.}

\end{figure}

\begin{figure}[ht!]
 \centering
 \begin{subfigure}[b]{0.49\textwidth}
  \centering
  \includegraphics[width=\textwidth,keepaspectratio=true]{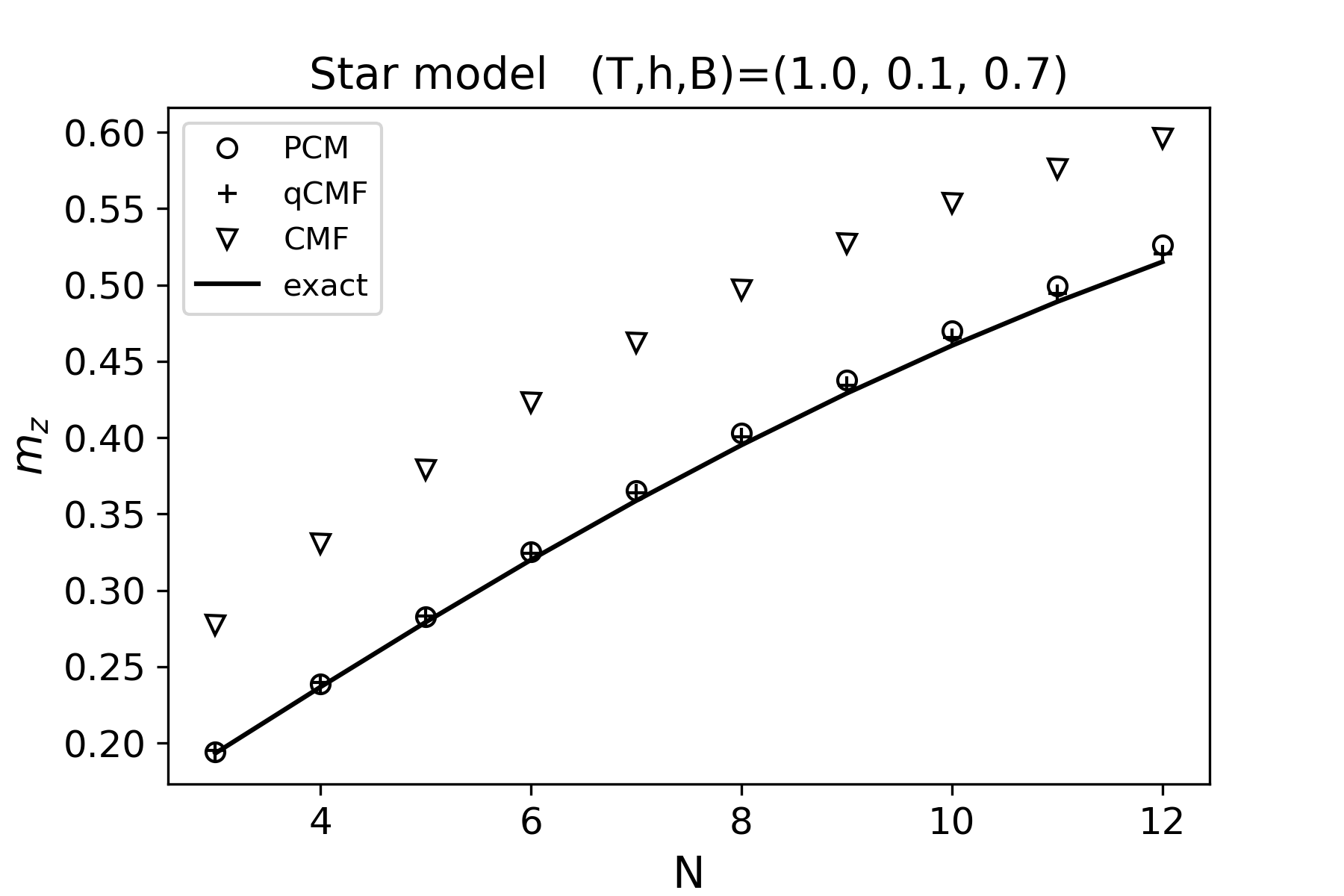}
 \caption{Longitudinal magnetization $m^z$.}
 \label{fig:star_model_mz}
 \end{subfigure} 
 \begin{subfigure}[b]{0.49\textwidth}
  \centering
  \includegraphics[width=\textwidth,keepaspectratio=true]{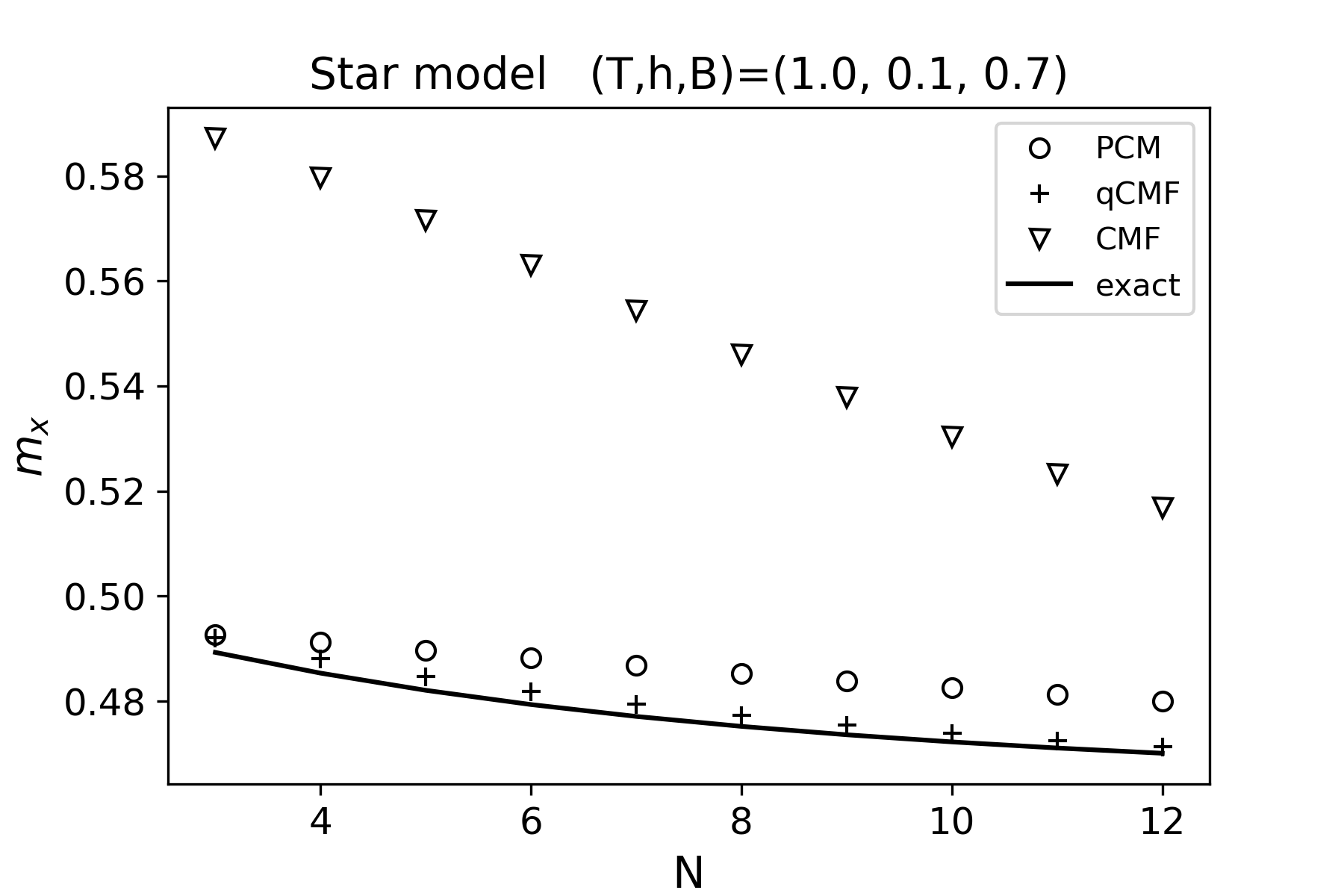}
 \caption{Transverse magnetization $m^x$.}
 \label{fig:star_model_mx}
 \end{subfigure} 
 \caption{\label{fig:star_model} Longitudinal and transverse magnetizations for a star-like model for different system sizes. The figure shows the results for the qCMF and PCM approximations for a spin in the border [see figure \eqref{fig:star_model_diagram}]. It also shows the prediction of the original CMF and the exact values (continuous line). The qCMF gives accurate results in both the longitudinal and transverse directions with a computational cost that scales linearly with the number of neighbors.}
\end{figure}

Our qCMF can be understood an upgrade of the cavity mean field of \cite{Dimitrova}, which in turn is a simplification of the PCM. Therefore, we want to clarify how our proposal compares with
PCM and establish whether it is able to recapture the desired accuracy while preserving a linear scaling of the computational complexity. To this end, we define a simple star-like lattice with a spin $i$ in the center and $N-1$ spins connected to it as in figure \eqref{fig:star_model_diagram}. We then compare the $m^z$ magnetizations predicted by the three approximations with the exact value for different connectivities. We use an external field $h_z = 0.1$  to break the z-symmetry of the Hamiltonian and $B = 0.7$, which is an intermediate
value between the two classical limits $B = 0$ and $B\gg 1$. The temperature is fixed to $T = 1.0$.
The interaction constant is $J = 1$. The results, shown in figure \eqref{fig:star_model}, are qualitatively similar for other values of the model parameters. 
The cavity distribution of a border spin, $\MUij{j}{i}$, is trivial and does not require any projection. In the absence of $i$ the border spin is isolated. As a consequence, the PCM statistics of the central spin  correspond trivially with the exact calculation. It is more informative to look
at one of the spins $j$ in the border and compare its observables when $\MUij{i}{j}$ (the cavity distribution of the central spin) is computed via PCM or qCMF. Figure (\ref{fig:star_model_mz}) shows that both algorithms accurately estimate the $m^z$ magnetization
for all system sizes. Calculations also show that qCMF gives better results than PCM for the transverse magnetization $m^x$, especially for high connectivities. This last finding is somewhat unexpected because PCM makes use of an exact diagonalization of the Hilbert subspace of spin $i$ and its neighbors. A plausible explanation is that PCM is fine-tuned to approximate $m^z$ through moment matching of the real and projected $\mu$ distributions, while $m^x$ remains uncontrolled. On the other hand, we have that qCMF explicitely separates the trajectories with no jumps from the rest before projecting the $m_z$ moment. In doing so, qCMF indirectly provides a partial account for the jump statistics. Moreover, since $m^x$ is proportional to the average number of jumps in a trajectory [see Appendix \eqref{app:observables}], we can expect that the qCMF approximation would have a positive impact on the $m_x$ estimation as well.
In figure (\ref{fig:star_model_mz}) we also display the results for the CMF algorithm (the one without the classical correction) which, as expected, gives poorer estimates than its counterparts. 

From this analysis we can conclude that a careful treatment of the balance between classical and quantum trajectories in the distributions $\MUij{i}{j}$ is enough to obtain a precise estimate of $m^z$ while reducing the computational cost from exponential to linear in the number of neighbors. Moreover, we observe that our approximation (qCMF) is able to outperform the more costly procedure of exact diagonalization for the transverse magnetization. 


\begin{figure}[ht]
\centering
  \includegraphics[width=\textwidth,keepaspectratio=true]{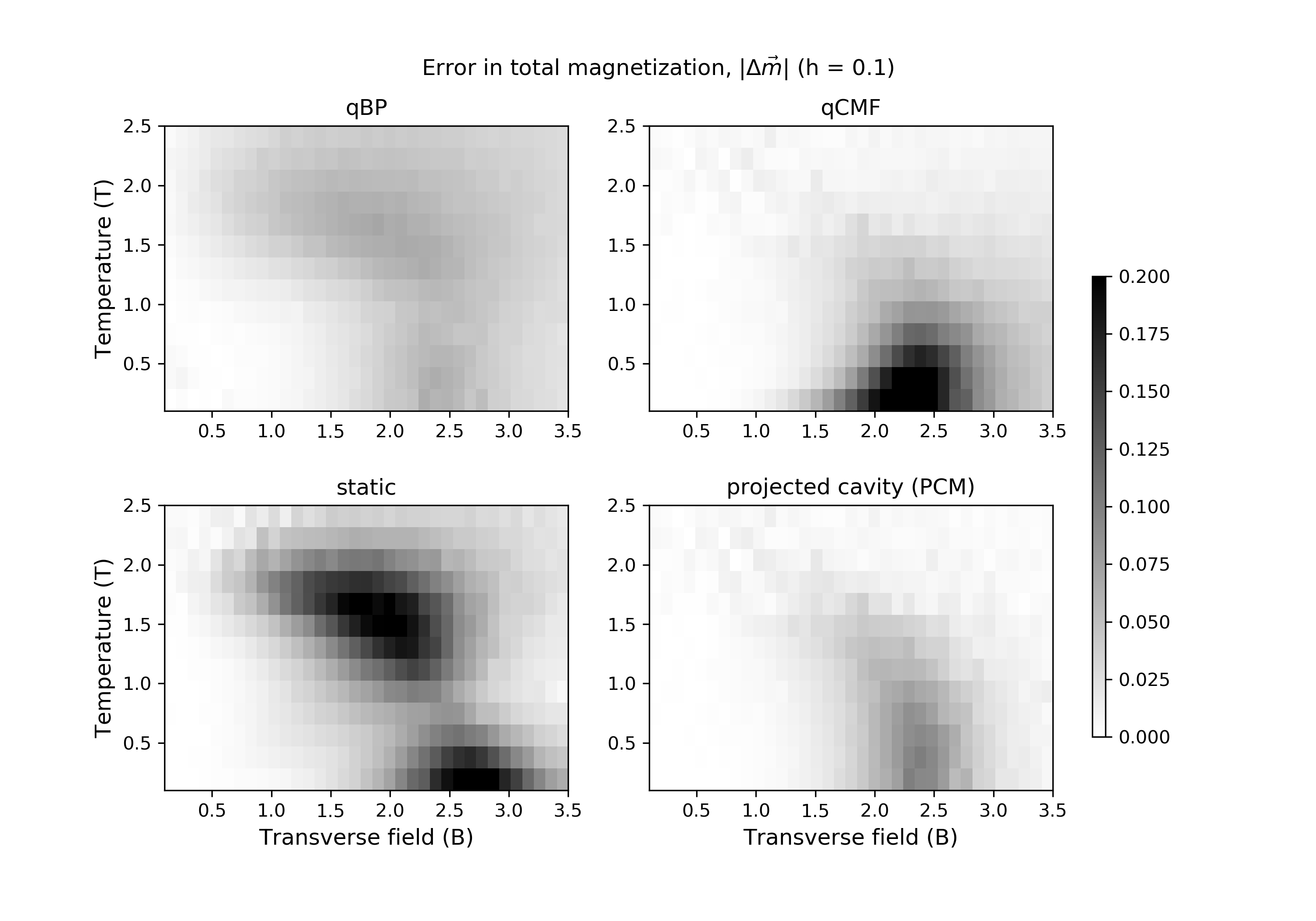}
 \caption{\label{fig:error_total_magnetization_infinite}Error in 
 single spin total magnetization $|\Delta \vec m|$ for several approximate inference algorithms. Each panel shows a heat map of the absolute value  of the vector difference $|\Delta \vec m| = |\vec m^{\mbox{\tiny approx}} - \vec m^{\mbox{\tiny ref}}|$ in the $B$ vs. $T$ phase plane. The reference value $\vec m^{\mbox{\tiny ref}}$ is
 computed from a numerical solution of the exact cavity equations  \cite{florent}. The static approximation gives the worst results whereas the PCM is the most precise overall. Far from the quantum phase transition ($T = 0, B_c \approx 2.232$), our qCMF is the more efficient alternative to make a precise inference. The PCM and qBP, by design focused on longitudinal $m^z$ estimation, are more suited for analysing the critical region where this quantity is the order parameter.}
\end{figure}

We now proceed to a more thorough benchmarking on another typical toy model:
an infinite tree (or random regular graph) of fixed connectivity and homogeneous ferromagnetic interactions. The connectivity of the graph is $c = 3$. We show in figure (\ref{fig:error_total_magnetization_infinite}) the $B$ vs. $T$ heatmap of deviations of the total magnetization $|\Delta \vec m| = |\vec m^{\mbox{\tiny approx}} - \vec m^{\mbox{\tiny ref}}|$ from the exact values for the SA, PCM, qCMF and qBP approximations. By comparing
the total magnetization we can get a better idea of the overall performance and make an assessment that includes the transverse magnetization as well. As a reference value $\vec m^{\mbox{\tiny ref}}$ we use the prediction of the cavity method in continuous time computed as in \cite{florent}. The first thing we notice is that all approximations suffer in the vicinity of the quantum phase transition, located at $B_c \approx 2.232$. Close to this region, the best results are obtained by the qBP and PCM. This is in line with the fact that these two approximations are designed to be accurate for $m^z$, the order parameter of this transition. Considering the overall performance, the best results correspond to PCM and the worst to the static approximation. In the regions away from the transition, we observe that the qCMF gives precise estimates, being in this regime faster than PCM and more accurate than qBP. The main source of error for qBP is in the estimate of the transverse magnetization $m^x$ (not shown).

\begin{figure}
\centering
  \includegraphics[width=1.0\textwidth,keepaspectratio=true]{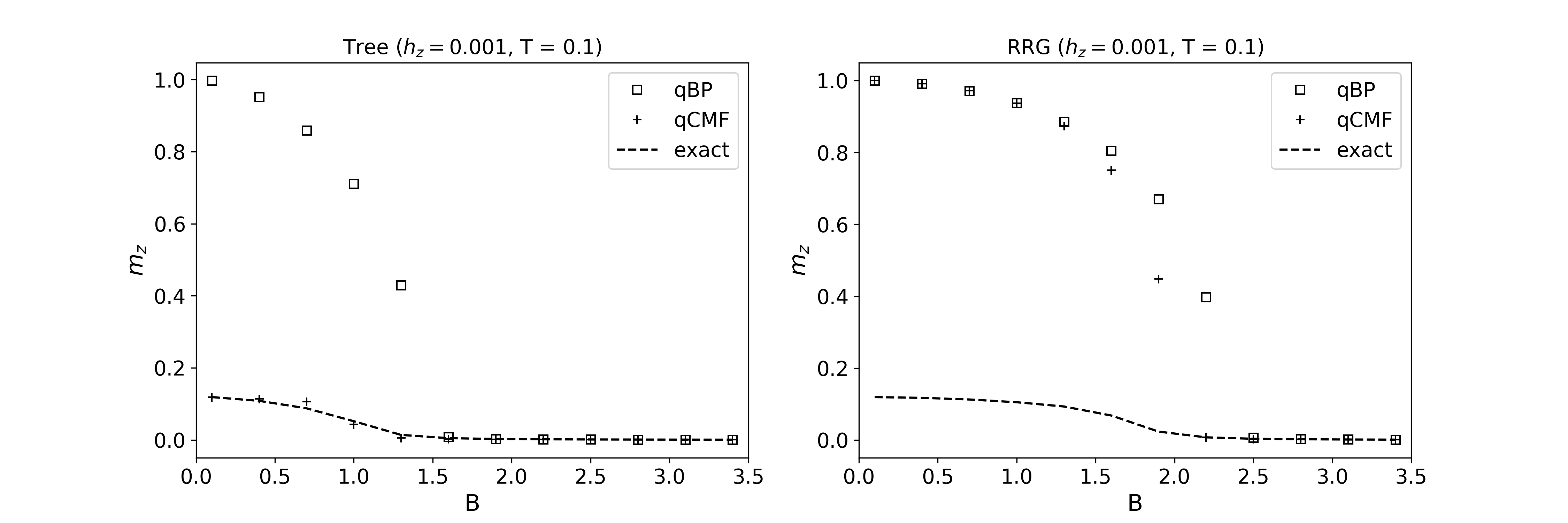}
 \caption{\label{fig:tree_rrg_loops}Comparing qBP and qCMF average $m^z$ magnetization for a ferromagnetic finite tree (left) and a random regular graph (right) as a function of the transverse field for low temperature. The simulation shows that for a tree, qBP is unable to find the correct solution and gets caught by the effects of induced loopiness. In contrast, qCMF works well in the tree, accounting for the balance between all spins up and down solutions. In the random regular graph, with physical loops, both algorithms fail to provide an accurate result. An increase of the longitudinal field $h_z$ would break the symmetry of the low temperature solution and the difference between the exact and approximate results would diminish.}
\end{figure}

The above results suggest that the qBP and qCMF algorithms are reasonable candidates for efficient approximate inference. The PCM is substantially better only near the phase transition. We now test both qBP and qCMF in finite networks to gain more insight on the effect of the system size and short loops. We consider two finite lattices of 12 spins each. One is a tree, where the connectivity of internal nodes was kept at $c = 3$ for as many nodes as possible, and the other is a random regular graph with $c = 3$. We show in figure \eqref{fig:tree_rrg_loops} the average longitudinal magnetization $m^z$ as a function of the transverse field $B$. 
In the plots shown, the temperature of the systems is low ($T = 0.1$) and the longitudinal external field $h_z = 0.001$ is very small, although not negligible. The exact calculation (dashed line) shows that for both topologies there is a nonzero $m^z$ magnetization. This polarization is small due to the finite system size. For many more spins the external $h_z$ would be enough to break the symmetry more evidently. Actually, in the thermodynamic limit, the state of the system corresponds to a ferromagnetic phase when $B < B_c = 2.232$. 

For the loopy random regular graph (RRG, right)
both qCMF and qBP converge to the (incorrect) infinite system ferromagnetic solution. This
is a consequence of the structure of the RRG. For local algorithms the topology is indistinguishable from the infinite case and the short loops makes them converge to the selfconsistent solution of the infinite lattice. For the finite tree (left) we have a different outcome. In this case there are no real loops and the qCMF succeedes in converging to the correct results. The qBP fails in this case, converging to a magnetized solution. This behavior is connected to the induced loopiness of qBP, explained in the previous section. The tree and the RRG are both loopy in what concerns to qBP. Another complication due to the loopy behaviour of qBP in a tree is that one cannot exploit this structure to 
speedup the convergence. With qCMF (as in the classical BP algorithm) one can schedule the computation of the distributions so that only one pass over the lattice is required. For qBP one has to perform a much slower fixed point iteration. These qualitative differences between the tree and the random regular graph are lost for larger values of $h_z$, for which the symmetry of the low temperature states is broken. In that regime both qBP and qCMF approximate well the exact value of the observables.

Summarizing, amongst the methods used to compute the approximate solution of the quantum cavity equations, qBP and the newly proposed qCMF stand out for two fundamental properties. First, both scale linearly with the connectivity of the network, which allows to consider them for more densely connected systems and second, both provide accurate estimates of observables for an ample range of physical conditions and different topologies. For transverse observables, our qCMF is more accurate. The qBP algorithm by construction cannot take full advantage of factorized tree structures.
\section{Conclusions}
\label{sec:conclusion}
In this work we have explored different approximate inference methods for the transverse field Ising model and proposed a new ansatz for the solution of the quantun cavity equations. We have shown that the quantum cavity formalism is a starting point for a wide range of approximations, including those in the preceding literature (qBP, PCM, naive mean field, CMF, static). Our proposal for a new algorithm, the quantum Cavity Mean Field, proved to be efficient and accurate in most regimes, with the largest errors in the vicinity of a phase transition. There are several advantages in the use of this method: it is consistent in the classical limit of zero transverse field, the computational cost scales linearly with the connectivity of the lattice, gives better estimates of transversal 
observables and preserves the local tree-like structure of the network (no induced loopiness). The method is based on a separation of the classical and quantum parts of the cavity distributions. We can speculate that this separation could be useful in the design of sampling strategies for quantum MC simulations given that classical paths are very cheap to sample.

Looking forward to applications, we expect qCMF to be a useful approximation for highly connected lattices, like
the ones found in learning problems, e.g. for quantum Boltzmann machines. Also, qCMF can contribute to theoretical approaches to disordered quantum systems. In this context, the advantage with respect to PCM and qBP is that a fixed point intermediate step, inconvenient when averaging over quenched disorder, is completely avoided. An extension of the algorithm presented here to Hamiltonians with higher order interactions in the Z direction should be straighforward. In this case, the distintion between $\mu$ and $M$ distributions is even more meaningful and the most convenient graphical
representation of the system structure is given by a factor graph \cite{yedidia}. An interesting direction for future work would be to
analyze other members of the stoquastic Hamiltonians family, for example,
those including $\paux_i \paux_j$ interaction terms. This is a harder
task since the ST expansion of the corresponding density matrix cannot be
decomposed in independent factors as for the TFIM.

\appendix

\section{Observables in the path integral formulation}
\label{app:observables}
Observables $\hat F$ that are diagonal in the computational basis of the eigenvectors of $\ket{\pauz}$ have a very simple form in the path integral formalism \cite{florent}. If the operator expression is $\hat F = f(\pauz_1,\ldots,\pauz_N)$ then the corresponding observable $F(\bfs)$ is:
\begin{equation}
F(\bfs) = \dfrac{1}{N_s} \sum_{\tau = 1}^{N_s} f(\s_1^\tau,\ldots,\s_N^\tau)
\label{eqn:trajectory_observable}
\end{equation}
and the average value $\langle \hat F \rangle = \tr{\r \hat F}$ is just $\langle \hat F \rangle = \sum_{\bfs} F(\bfs) \rho(\bfs)$. The index $\tau$ labels the Suzuki-Trotter slices. These formulas have obvious generalizations in the $N_s\rightarrow \infty$ limit.

Often needed observables are the spin magnetization and the correlations. For the Z magnetization
we have $\hat F \equiv \pauz_i$. Then, according to \eqref{eqn:trajectory_observable}:
$$
F(\bfs) = m(\vec\s_i) = \dfrac{1}{N_s} \sum_{\tau = 1}^{N_s} \s_i^\tau
$$
and marginalizing the joint probability distribution $\rho(\bfs)$ we get:
\begin{equation}
 m^z_i  = \sum_{\vec\s_i} m(\vec\s_i) \rho_i(\vec\s_i)  
\end{equation}
In the same way, for the nearest neighbor correlation in the Z direction we obtain:
\begin{equation}
 c^{zz}_{ij} =  \sum_{\vec\s_i,\vec\s_j} \vec\s_i\cdot\vec\s_j\; \rho_{ij}(\vec\s_i, \vec\s_j) 
\end{equation}
where $\vec\s_i\cdot\vec\s_j = \dfrac{1}{N_s} \sum_{\tau = 1}^{N_s} \s_i^\tau \s_j^\tau$, as defined also
in the main text.

Transversal observables are not so straightforward. The single spin $m^x_i = \expectation{\paux_i}$ magnetization is related to the average of the number of flips $n_f(\vec\s_i)$ in the trajectory $\vec\s_i$:
\begin{equation}
 m^x_i = \dfrac{1}{\beta B_i}\sum_{\vec\s_i} n_f(\vec\s_i) \rho_i(\vec\s_i)
\end{equation}
Likewise, the transversal correlation $c_{ij}^{xx}$ corresponds to the average number of simultaneous jumps
in trajectories $\vec\s_i$ and $\vec\s_j$, denoted $n_f^{ij}(\vec\s_i,\vec\s_j)$:
\begin{equation}
 c_{ij}^{xx} = \dfrac{1}{\beta^2  B_i B_j} \sum_{\vec\s_i,\vec\s_j} n_f^{ij}(\vec\s_i,\vec\s_j) \rho_{ij}(\vec\s_i,\vec\s_j)
\end{equation}

\section{Details of the qCMF algorithm}\label{app:details_qCMF}
The crucial point in the qCMF Algorithm \eqref{alg:qCMF} is the computation of the parameters
$c_{ij}(s)$ and $m^q_{i\rightarrow j}$ once we plug in the $\Mij{k}{i}$ distributions:
\begin{equation}
 \Mij{k}{i} =  q_{ki} \exp\left[\beta J_{ik} m(\vec\s_i) m_{k\rightarrow i}^q \right] + \sum_{s=\pm 1} c_{ki}(s) \exp\left[\beta J_{ik} m(\vec\s_i)s\right]
 \label{eqn:M_update_appendix}
\end{equation}
on the RHS of equation \eqref{eqn:message_passing_mu}.
The functions $\Mij{k}{i}$ depend only on $m(\vec\s_i)$, the average value in the imaginary time direction. This allows us to sum in equation \eqref{eqn:message_passing_mu} over all
trajectories $\vec\s_i$ that have the same value of $m(\vec\s_i)=m$:
\begin{align}
 \mu_{i\rightarrow j}(m) &= \sum_{\vec\s_i:m(\vec\s_i)=m} \MUij{i}{j} \\
 &= \dfrac {1}{Z^\mu_{i\rightarrow j}}w_s(m,B_i)\exp[\b h_i m] \prod_{k\in \partial i\setminus j} M_{k\rightarrow i}(m)
 \label{eqn:mu_m}
\end{align}
where $Z^\mu_{i\rightarrow j}$ is a normalization fixed by the condition $\int\mu_{i\rightarrow j}(m)  dm = 1$ and:
\begin{align}
 w_s(m,B_i) &= \sum_{\vec\s_i:m(\vec\s_i)=m} w(\vec\s_i,B_i) \\
 &= \delta(m-1) + \delta(m+1) + \dfrac{\b B_i}{\sqrt{1-m^2}} I_1(\b B_i\sqrt{1-m^2}).
 \label{eqn:w_m}
\end{align}
Here $I_1$ is the modified Bessel function of the first kind.
For a derivation of \eqref{eqn:w_m} the reader can check the appendices in \cite{florent}.
In the structure of $w_s(m,B_i)$ we recognize the separation of classical $(m=\pm 1)$ and quantum trajectories. We can readily identify the coefficients $c_{ij}(s)$:
\begin{equation}
 c_{ij}(s) = \dfrac{\exp[\b h_i s]}{Z^\mu_{i\rightarrow j}} \prod_{k\in \partial i\setminus j} M_{k\rightarrow i}(s)\;\;\;\; s = \pm 1
 \label{eqn:classical_weights}
\end{equation}
The average magnetization $m^q_{i\rightarrow j}$ is computed by integrating the regular part
of $\mu_{i\rightarrow j}(m)$:
\begin{equation}
 m^q_{i\rightarrow j} = \dfrac{1}{q_{ij}Z^\mu_{i\rightarrow j}}\int_{-1}^1 m\left[\dfrac{\b B_i}{\sqrt{1-m^2}} I_1(\b B_i\sqrt{1-m^2})\exp[\b h_i m] \prod_{k\in \partial i\setminus j} M_{k\rightarrow i}(m)\right]dm
\end{equation}
As a stated in the main text, $q_{ij} = 1 - \sum_{s=\pm 1} c_{ij}(s)$ is a rescaling factor
that normalizes the quantum (regular) part of $\mu_{i\rightarrow j}(m)$.

Once the algorithm \eqref{alg:qCMF} has converged, it is possible to compute the estimate of the  real magnetization $m_i^z = \langle m(\vec\s_i)\rangle_{\rho_i}$ after noticing that the trajectory
distribution $\rho_i(\vec\s_i)$ differs from $\MUij{i}{j}$ only in a factor $M_{j\rightarrow i}(\vec\s_i)$. Therefore, it can also be written as $\rho_i(m(\vec\s_i))$ and the previous simplifications are valid. The final formula is:
\begin{align}
\nonumber
 m_i^z &= \dfrac{1}{Z_i}\left[\sum_{s=\pm1}s\exp[\b h_i s] \prod_{k\in \partial i} M_{k\rightarrow i}(s)\right.\\
 &+ \left.\int_{-1}^1 m\left(\dfrac{\b B_i}{\sqrt{1-m^2}} I_1(\b B_i\sqrt{1-m^2})\exp[\b h_i m] \prod_{k\in \partial i} M_{k\rightarrow i}(m)\right)dm\right]
\end{align}
with $Z_i$ being the normalization constant of the distribution $\rho_i(\vec\s_i)$.
To  compute $m_i^x$ we use the relation $m_i^x = \frac{1}{\b}\frac{\partial \ln Z_{i}}{\partial (B_i)}$ to get:
\begin{equation}
 m_i^x = \dfrac{1}{Z_i}\int_{-1}^1\left[ \b B_i I_0(\b B_i\sqrt{1-m^2})\exp[\b h_i m] \prod_{k\in \partial i} M_{k\rightarrow i}(m)\right] dm
\end{equation}

\subsection{Bethe free energy estimation}
\label{subsec:bethe_free_energy}
In this section we sketch the derivation of the exact cavity equations and show how to compute
an approximated value of the Bethe free energy by using the solution provided by the qCMF algorithm. 
The standard procedure we follow \cite{yedidia} is to minimize the Bethe free energy of the classical system \eqref{eqn:ST_density} obtained after the Trotterization. 

The Bethe free energy $F_{\text{B}}$ is a region-based approximation of the exact free energy. In general, it is a functional of local probability distributions and for a system with pairwise interactions it is composed of two different parts:
\begin{equation}
 F_{\text{B}}[\{\rho_i, \rho_{ij}\}] = \sum_{(ij)} F_{ij}[\rho_{ij}] + \sum_i (1- d_i) F_i[\rho_i]
 \label{eqn:bethe_free_energy}
\end{equation}
The first sum in \eqref{eqn:bethe_free_energy} contains the free energy contribution $F_{ij}$ of all interacting pairs. The second part accounts for all single-spin free energies, $F_i$. The $(1-d_i)$ factors, with $d_i$ the degree of spin $i$ (number of neighbors), balance the overcounting effect due to ovelaps between regions. $\rho_i$ and $\rho_{ij}$ are the local trajectory distributions.

The free energy of a region $R$ (pair or single site) is the usual sum of an average energy and an entropy part:
\begin{equation}
 F_{R} = E_{R} - \dfrac{1}{\b} S_{R}
\end{equation}
with $E_R = \sum_{\vec\s_R} E_R(\vec\s_R) \rho_R(\vec\s_R)$ and $S_R = -\sum_{\vec\s_R} \rho_R(\vec\s_R) \ln \rho_R(\vec\s_R)$.

In our case, the relevant local energy functions are:
\begin{align}
 E_i(\vec\s_i) &= - h_i m(\vec\s_i) - \dfrac{1}{\beta}\ln w(\vec\s_i,B_i) \\
 E_{ij}(\vec\s_i,\vec\s_i) &= E_i(\vec\s_i) + E_j(\vec\s_j) - J_{ij} \vec\s_i\cdot\vec\s_j
\end{align}

The key step is to minimize \eqref{eqn:bethe_free_energy} with respect to the $\rho_R(\vec\s_R)$ distributions while taking care of the consistency among them. For the Bethe approximation the typical
constraint is that pair distributions should marginalize to single site ones:
\begin{equation}
 \rho(\vec\s_i) = \sum_{\vec\s_j} \rho_{ij}(\vec\s_i,\vec\s_j)
 \label{eqn:constraint}
\end{equation}
This is a constrained minimization problem that is easily solved by adding some Lagrange multipliers.
After some algebra, it can be shown \cite{yedidia} that the local probabilities will have the form:
\begin{equation}
 \rho_i(\vec \s_i) = \dfrac{1}{Z_i} w(\vec \s_i,B_i)
 \exp\left[\beta h_i m(\vec \s_i)\right] 
 \prod_{k \in \partial i} \Mij{k}{i}
 \label{eqn:single_site_distribution_M}
\end{equation}
\begin{equation}
 \rho_{ij}(\vec \s_i,\vec \s_j) = \dfrac{1}{Z_{ij}} 
 \exp\left[\beta (J_{ij} \vec\s_i\cdot\vec\s_j + h_i m(\vec \s_i)+ h_j m(\vec \s_j))\right] w(\vec \s_i,B_i)w(\vec \s_j,B_j)\prod_{k \in \partial i\setminus j} \Mij{k}{i}\prod_{k' \in \partial j\setminus i} \Mij{k'}{j}
 \label{eqn:pair_distribution_M}
\end{equation}
where $\Mij{k}{i}$ are simple functions of the Lagrange multipliers. The constraints \eqref{eqn:constraint} give the self-consistent equations for the $\Mij{k}{i}$. See equation \eqref{eqn:message_passing_mu_no_mu}. One can finally define $\MUij{i}{j}$ as in \eqref{eqn:message_passing_mu}.

The Bethe free energy functional evaluated in the solution given by \eqref{eqn:single_site_distribution_M} and \eqref{eqn:pair_distribution_M} has a neat expression
in terms of the normalization constants $Z_i, Z_{ij}$ \cite{Rizzo2010}:
\begin{equation}
 -\beta F_{\text{B}} = \sum_{(ij)} \ln Z_{ij} + \sum_{i} (1- d_i) \ln Z_i
\end{equation}
We can now employ the qCMF solution to compute an approximation for these local partition functions, using the fact that $M$ distributions depend only on $m(\vec\s_i)$:
\begin{align}
 Z_i &\approx \int_{-1}^1 dm\; \exp\left[\beta h_i m\right] w_s(m,B_i) \prod_{k \in \partial i} M_{k\rightarrow i} (m) \\
 Z_{ij} &\approx \int_{-1}^1 dm dm' \; \exp\left[\beta (J_{ij} m m' + h_i m + h_j m')\right] w_s(m,B_i) w_s(m',B_j) \prod_{k \in \partial i
 \setminus j} M_{k\rightarrow i} (m)\prod_{k' \in \partial j
 \setminus i} M_{k'\rightarrow j} (m')
\end{align}
For $Z_{ij}$ we had to further approximate $\vec\s_i\cdot\vec\s_j$ as $mm'$.

\subsection{Classical limit of qCMF}
\label{subsec:classical_limit}
In this subsection we show that in the classical limit qCMF recovers the exact solution for the cavity distributions. To keep the discussion self-contained, we first summarize the classical cavity results.

In the limit $B\rightarrow 0$ the eigenvalues of the TFIM Hamiltonian \eqref{eqn:ising_hamiltonian} reduce to a classical Ising energy function: 
$E(\{s\}) = - \sum_{(ij)} J_{ij} s_i s_j - \sum_{i} h_i s_i$, if one works in the computational basis $\ket{s_1\ldots s_N}$ \footnote{We have used $s$ instead of $\s$ in this section to avoid the risk
of confusion with trajectory variables.}. The density
operator is diagonal, with entries corresponding to the usual Boltzmann distribution $\rho(\{s\})\propto \exp-\b E(\{s\})$. When the topology is tree-like, as it is the case in the present paper, the single variable $\rho_i(s_i)$ and pair $\rho_{ij}(s_i,s_j)$ marginals of the joint probability are provided by the well-known BP algorithm 
\cite{yedidia}. The local probabilities $\rho_i,\rho_{ij}$ can be written in terms of \textit{classical} cavity distributions $M',\mu'$ using equations that are almost identical to \eqref{eqn:single_site_distribution} and \eqref{eqn:pair_distribution}:  
\begin{equation}
 \rho_i(s_i) = \dfrac{1}{Z_i}
 \exp\left[\beta h_i s_i\right] 
 \prod_{k \in \partial i} M'_{k\rightarrow i}(s_i)
 \label{eqn:single_site_distribution_classical}
\end{equation}
\begin{equation}
 \rho_{ij}(s_i,s_j) = \dfrac{1}{Z_{ij}} 
 \exp\left[\beta J_{ij} s_i s_j \right] \mu'_{i\rightarrow j}(s_i)  \mu'_{j\rightarrow i}(s_j)
 \label{eqn:pair_distribution_classical}
\end{equation}
the main differences being that the factor $w_i$ is not present in \eqref{eqn:single_site_distribution_classical} and that $s_i,s_j$ are discrete $\pm1$ variables rather than trajectories. The cavity distributions are computed as the fixed point of the following set of equations:
\begin{eqnarray}
\label{eqn:message_passing_mu_classical}
 \mu'_{i\rightarrow j}(s_i) &=& \dfrac{1}{Z_{i\rightarrow j}^\mu} \exp\left[\beta h_i s_i\right] 
 \prod_{k \in \partial i\setminus j} M'_{k\rightarrow i}(s_i)\\ 
 M'_{k\rightarrow i}(s_i) &=& \dfrac{1}{Z_{k\rightarrow i}^M} 
 \sum_{s_k = \pm 1}\exp\left[\beta J_{ik} s_i s_k \right]\mu'_{k\rightarrow i}(s_k)
 \label{eqn:message_passing_M_classical}
\end{eqnarray}
We will now show that when $B\rightarrow 0$, the $M$ and $\mu$ distributions obtained by the qCMF algorithm satisfy \eqref{eqn:message_passing_mu_classical} and \eqref{eqn:message_passing_M_classical} and therefore provide the correct solution to the inference problem in this case.

The starting point of the argument is to note that the quantum weight \eqref{eqn:ST_w} will go to zero for trajectories with jumps when $B\rightarrow 0$. As a consequence, $w_s(m,B \rightarrow 0)$ in \eqref{eqn:w_m} will consist only of the two delta function terms and all the cavity distributions will concentrate on the classical trajectories $\vec \s \equiv  s = \pm 1$. Furthermore, $\mu_{i\rightarrow j}(m)$ as defined in \eqref{eqn:mu_m}, will be supported only for $m = \pm 1$. We have already called $c_{ij}(s)$ these classical weights [see below \eqref{eqn:quantum_classical_separation_MU}]. However, again for convenience, we will denote them as $\mu^c_{i\rightarrow j}(s_i)$. Equation \eqref{eqn:classical_weights} provides the update rule:
\begin{equation}
 \mu^c_{i\rightarrow j}(s_i) = \dfrac{\exp[\b h_i s_i]}{Z^\mu_{i\rightarrow j}} \prod_{k\in \partial i\setminus j} M_{k\rightarrow i}(s_i)
 \label{eqn:classical_update_mu}
\end{equation}
Due to the concentration of the probability on the classical trajectories, the quantum weights $q_{ki}$ become zero
and there will be no need to compute the first term of \eqref{eqn:M_update_appendix}. This last equation now reads:
\begin{equation}
 \Mij{k}{i} =  \sum_{s_k=\pm 1} \mu^c_{k\rightarrow i}(s_k) \exp\left[\beta J_{ik} m(\vec\s_i)s_k\right]
 \label{eqn:M_update_appendix_2}
\end{equation}
Since we only need $\Mij{k}{i}$ for $\vec \s_i \equiv s_i = \pm 1$ and the average magnetization is $m(\vec \s_i) = s_i$ for such trajectories, we can very well restrict \eqref{eqn:M_update_appendix_2} to:
\begin{equation}
 M_{k\rightarrow i}(s_i) =  \sum_{s_k=\pm 1} \mu^c_{k\rightarrow i}(s_k) \exp\left[\beta J_{ik} s_i s_k\right]
 \label{eqn:M_update_appendix_3}
\end{equation}

So, we have shown that the relevant weight of the cavity distribution concentrate on the classical trajectories
for $B\rightarrow 0$ and the update equations \eqref{eqn:classical_update_mu} and \eqref{eqn:M_update_appendix_3}
are identical to the classical ones, \eqref{eqn:message_passing_mu_classical} and \eqref{eqn:message_passing_M_classical}. The reader can check that local distributions \eqref{eqn:single_site_distribution} and \eqref{eqn:pair_distribution} display this concentration of the probability as well.

\bibliographystyle{unsrt}
\bibliography{general}

\end{document}